\documentclass{elsart3p}

\usepackage{graphicx}

\usepackage{amssymb}

\begin{document}

\begin{frontmatter}



\title{Magnetic response of noncentrosymmetric superconductor La$_2$C$_3$ :
Effect of double-gap and spin-orbit interaction}


\author[lab1,lab2]{R. Kadono\corauthref{cor}}
\corauth[cor]{Tel. +81-29-864-5625,
Fax: +81-29-864-5623,
email: ryosuke.kadono@kek.jp}
\author[lab2]{M. Hiraishi}
\author[lab2]{M. Miyazaki}
\author[lab2]{K. H. Satoh}
\author[lab1]{S. Takeshita}
\author[lab3]{S. Kuroiwa}
\author[lab3]{S. Saura}
\and \author[lab3]{J. Akimitsu}

\address[lab1]{Muon Science Laboratory, Institute for Materials Structure Science, High Energy Accelerator Research Organization (KEK), Tsukuba, Ibaraki 305-0801, Japan}
\address[lab2]{Department of Materials Structure Science, The Graduate University for Advanced Studies, Tsukuba, Ibaraki 305-0801, Japan}
\address[lab3]{Department of Physics, Aoyama Gakuin University, Sagamihara, Kanagawa 229-8558, Japan}

\begin{abstract}
The presence of spin-orbit (SO) interaction in a noncentrosymmetric superconductor, La$_2$C$_3$ ($T_c\simeq 11$ K) is demonstrated by muon spin rotation ($\mu$SR) in its  {\sl normal state}, where $\mu$SR spectra exhibit  field-induced weak depolarization due to van Vleck-like local susceptibility. In the mixed state, muon spin relaxation due to inhomogeneity of internal field ($\sigma_{\rm v}$) exhibits a field dependence that is characterized by a kink, where $\sigma_{\rm v}$ (and hence the superfluid density) is more strongly reduced at lower fields.  This is perfectly in line with the presence of a secondary energy gap previously inferred from the temperature dependence of $\sigma_{\rm v}$, and also consistent with the possible influence of asymmetric deformation of the Fermi surface due to the SO interaction.

\end{abstract}

\begin{keyword}
superconductivity \sep spin-orbit interaction \sep quasiparticle excitation

\end{keyword}
\end{frontmatter}


Multigap superconductivity is interesting in its own right as a manifestation of 
anisotropic superconductivity. 
Even within the framework of the BCS mechanism,  anisotropy in crystal structure
may lead to multi-band structure and associated 
complex superconducting order parameter because of the possible
coupling of electrons to different phonon modes upon formation of 
the Cooper pairs in the respective
energy bands.  The discovery of high-$T_c$ superconductivity 
in magnesium diboride (MgB$_2$)\cite{Nagamatsu:01} and subsequent struggle for 
proper understanding of its double-gap nature associated with 
$\sigma$- and $\pi$-bands brought the multigap 
superconductivity to the forefront of considerable attention in this field.

Recently, we have shown in a pair of sesquicarbide superconductors $Ln_2$C$_3$ ($Ln$=La, Y)
that the multigap structure would take great variety in its appearance, where, despite a common
double-gap structure in the order parameter, the temperature
dependence of superfluid density exhibits a remarkable difference  between La$_2$C$_3$ and Y$_2$C$_3$ that is understood as resulting from the alteration of 
interband coupling upon substitution of La for Y.\cite{Kuroiwa:08} 

Meanwhile, sesquicarbides are drawing further attention as a possible stage for exotic 
superconductivity caused by their noncentrosymmetric crystal structure.\cite{Fujimoto:07} 
The absence of inversion symmetry leads to mixing of parity in the Cooper pairs,
making it irrelevant to classify the states in terms of spin-multiplet.  It also gives rise to 
the spin-orbit (SO) interaction that may affect superconductivity as well as the electronic 
property of normal (paramagnetic) state. In particular, $Ln_2$C$_3$ 
may be in a unique situation that the SO interaction is of comparable magnitude to that
of superconducting gap, so that pair correlation between subbands (split by the SO
interaction) induced by external field 
might lead to a drastic change in the low energy properties of superconductivity. 
More specifically, considering the Dresselhaus-type interaction appropriate for
the crystal symmetry of $I\overline{4}3d$, the energy gap along the Fermi momentum  
${\bf k}_F\parallel (001)$, $(111)$, and $(100)$, for example, is 
$\sqrt{(\mu_BH_z)^2+\Delta^2}-\mu_BH_z$, so that it may be reduced to zero 
when $\mu_BH_z\gg\Delta$ [where $\Delta$ is the gap energy at zero field, $H_z$ is
the external field applied along $(001)$ axis].\cite{Fujimoto:07-2}  This means the occurrence of field-induced 
point nodes that would lead to enhanced quasiparticle excitation. 
It must be noted, however, that the presence of double-gap 
in $Ln_2$C$_3$ may require a more careful examination for this effect, since the 
suppression of smaller gap (having a smaller upper critical field $H_{c2}$) by external
field would show up in a similar way.  Here, we report on the result of field-induced effect
in La$_2$C$_3$ studied by muon spin rotation ($\mu$SR) experiment. 
\begin{figure}[tb]
\centering
\includegraphics[width=0.95\columnwidth]{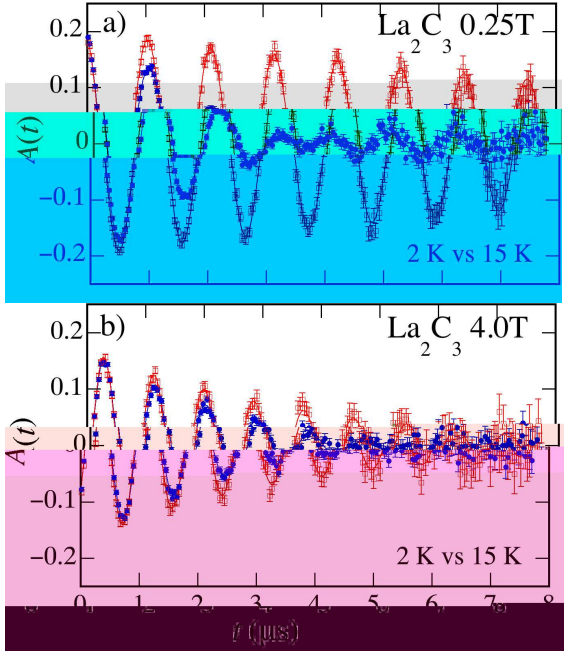}
\caption{Examples of TF-$\mu$SR time spectra displayed in 
the rotating reference frame of (a) 33 MHz and (b) 541 MHz, respectively.
In the paramagnetic state (15 K, shown by open symbols), the spectrum of 4.0 T exhibits 
faster relaxation than that of 0.25 T, indicating occurrence of field-indiced
magnetism. Additional relaxation due to the formation of flux line lattice
at 2 K (spectra shown in filled symbols) is seen in both cases. Solid curves
are fits by a Gaussian relaxation.
}\label{tspec}
\end{figure}

A conventional  $\mu$SR experiment was carried out
for a La$_2$C$_3$ sample ($T_c=10.9$ K) on the M15 beamline of TRIUMF, Canada, where 
details on the experiment are described in the previous report.\cite{Kuroiwa:08} 
The  sample  was common to the previous 
experiment, which turned out to be a typical double-gap superconductor 
[$\Delta_1(0) = 2.7(1)$ meV,  $\Delta_2(0) = 0.6(1)$ meV]. For the field-scan
measurements at 2 K (where both gaps 
are present), the sample was cooled down to the target temperature after the external
magnetic field was stabilized at a temperature above $T_c$ in order to minimize
the effect of flux pinning. Fig.~\ref{tspec} shows some examples of $\mu$SR time spectra 
obtained under a field of 0.25 and 4.0 T, where open symbols show the data in the 
normal state.  While the enhancement of spin relaxation upon cooling down 
to 2 K is observed for both cases (filled symbols), it is also noticeable that the 
relaxation {\sl in the normal state} is enhanced
by increasing external field.  Considering that La$_2$C$_3$ has a cubic structure 
(bcc, $I\overline{4}3d$) and that it is free from any local $d$ or $f$ electrons, we can attribute
this field-induced magnetization uniquely to the van Vleck-like paramagnetism
due to the SO interaction.  

\begin{figure}[tb]
\centering
\includegraphics[width=1.0\columnwidth]{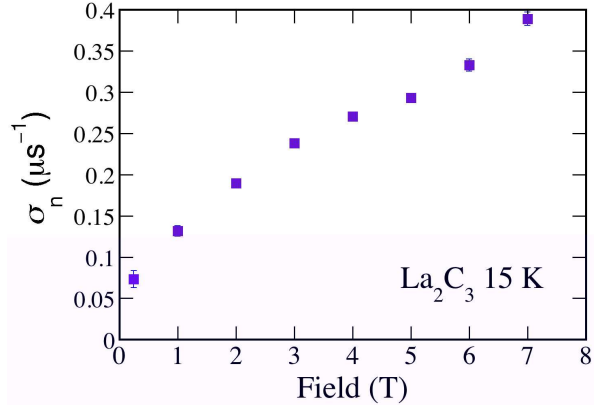}
\caption{Muon spin relaxation rate versus transverse field
observed in the normal state of La$_2$C$_3$ (at 15 K).
}\label{sn}
\end{figure}

The spin relaxation rate in the normal
state is deduced from fits using a Gaussian damping 
\begin{equation}
A(t)=A_0\exp(-\sigma_{\rm n}^2t^2)\cos(\gamma_\mu Bt+\phi), \label{gauss}
\end{equation}
where $A_0$ is the initial asymmetry, $\gamma_\mu$ is the muon gyromagnetic ratio (=135.53 MHz/T), 
$B$ is the local field felt by muons ($\simeq\mu_0H$, with $H$ being the external field), and $\phi$ is 
the initial phase of rotation. Although the fit with Eq.~(\ref{gauss}) does not reproduce
data at higher fields (particularly above 3 T) where the spectra exhibit a tendency toward
an exponential damping, we resort to this simple form for the convenience of 
evaluating qualitative trend for $\sigma_{\rm n}$ versus field.  As shown in Fig.~\ref{sn}, $\sigma_{\rm n}$
exhibits a quasi-linear dependence on the external field.  This is understood by considering
the field-induced van Vleck-like susceptibility ($\chi^{\rm vv}$) whose magnitude depends
on the direction of the primary axis for the SO interaction, so that it may lead to an inhomogeneity 
of effective field in polycrystalline sample, 
\begin{equation}
\sigma_{\rm n}^2\simeq\gamma_\mu^2 H^2\langle (\chi^{\rm vv})^2\rangle+\sigma_0^2,
\end{equation}
where $\sigma_0$ is the
contribution of random local fields from $^{139}$La nuclear magnetic moments.
(More specifically, $\chi^{\rm vv}$ also includes the Pauli paramagnetic term from the SO subbands.)
The slight change of $\sigma_{\rm n}$ observed around 3--4 T might be an artifact due to the
deteriorated quality of fit with Gaussian damping.  We also note that $\sigma_{\rm n}$ is mostly 
independent of temperature between 11 K and $\sim150$ K  under a field of 0.25 T.

\begin{figure}[tb]
\centering
\includegraphics[width=1.0\columnwidth]{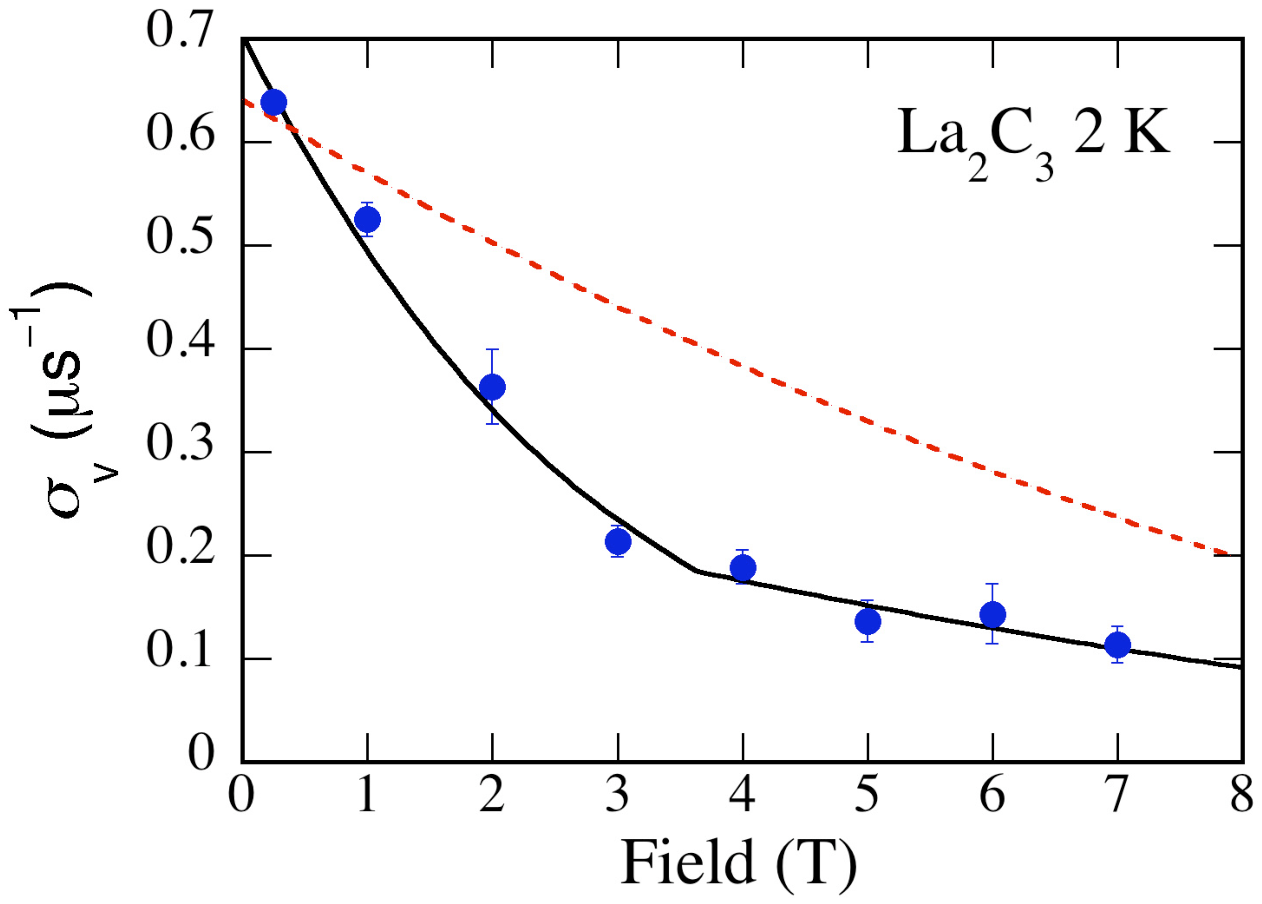}
\caption{Muon spin relaxation rate versus transverse field
observed in the mixed state of La$_2$C$_3$ (at 2 K). Dashed
curve shows $\sigma_{\rm v}$ expected for the case of single-gap ($s$-wave),
whereas solid curve is a fit using the double-gap model (see text).
}\label{sgmv}
\end{figure}

Considering the contribution of field inhomogeneity due to the van Vleck-like paramagnetism, 
we deduce the spin relaxation rate in the mixed state by fits of $\mu$SR time spectra 
using a form,
\begin{equation}
A(t)=A_0 G^{\rm n}(t;B)\exp\left[-\frac{1}{2}\sigma_{\rm v}^2t^2\right]\cos(\gamma_\mu Bt+\phi),
\end{equation}
where, instead of  eq.(\ref{gauss}), $G^{\rm n}(t;B)$ is chosen to best reproduce the spectra at 15 K ($>T_c$) at each
field. [$G^{\rm n}(t;B)$ consists of a sum of two exponential damping signals, where the parameters
describing $G^{\rm n}(t;B)$ is determined by the spectra at 15 K at each field and then fixed to these values
for the fit of spectra at 2 K to extract $\sigma_{\rm v}$ reliably.]
  In this definition, $\sigma_{\rm v}$ corresponds to the second moment for the 
field distribution [$B({\bf r})$] in the mixed state,\cite{Brandt:88}
\begin{equation}
\sigma_{\rm v}^2=\gamma_\mu^2\langle (B({\bf r}) -\mu_0H)^2\rangle.
\end{equation}
Fig.~\ref{sgmv} shows the deduced value of $\sigma_{\rm v}$ at respective
fields, where one can clearly observe a trend of steeper 
reduction with increasing field at lower fields (below $\sim$3 T).  

In the limit of extreme type II superconductors [i.e.,
$\lambda/\xi\gg1$, where $\lambda$ is the effective London penetration
depth and $\xi=\sqrt{\Phi_0/(2\pi H_{\rm c2})}$ is the Ginzburg-Landau
coherence length, $\Phi_0$ is the flux quantum, and $H_{\rm c2}$ is the
upper critical field], $\sigma_{\rm v}$ is determined by $\lambda$ using
a relation,\cite{Brandt:88}
\begin{equation}
\sigma_{\rm v}(h)=0.0274\frac{\gamma_\mu\Phi_0}{\lambda^2}(1-h)\sqrt{1+3.9(1-h)^2}\label{lmdh}
\end{equation}
where $h$ is the field normalized by the upper critical field ($h=H/H_{\rm c2}$). 
Although Eq.~(\ref{lmdh}) exhibits a tendency of concave curve, it does not
reproduce the field dependence of $\sigma_{\rm v}$ observed in Fig.~\ref{sgmv}.
This is particularly true when the known value of $H_{\rm c2}$ ($\simeq13$ T
at 2 K) is considered; the situation is illustrated by a dashed curve in Fig.~\ref{sgmv}.

Since La$_2$C$_3$ is known to have a double-gap structure in the order parameter,
it is natural to expand Eq.~(\ref{lmdh}) into the following form,
\begin{equation}
\sigma_{\rm v}(h) = w\sigma_{\rm v}(h_1)+(1-w)\sigma_{\rm v}(h_2),\label{lmd2h}
\end{equation}
where $h_i=H/H_{c2}^{(i)}$, and $H_{c2}^{(i)}$ is the upper critical field corresponding to
the respective energy gap [$\sigma_{\rm v}(h_i)$ must be set to zero
for $h_i>1$].  The linear combination of two components is appropriate, as $\sigma_{\rm v}$
is proportional to the superfluid density (i.e., $\sigma_{\rm v}\propto n_{\rm s}$). 
The solid curve in Fig.~\ref{sgmv} is the best fit with Eq.~(\ref{lmd2h}),
which yields $\lambda=390(3)$ nm,  $H_{c2}^{(1)}=16(4)$ T, $H_{c2}^{(2)}=3.6(4)$ T, 
and $w=0.41(9)$.  

It is interesting to note that the ratio of the upper critical field [$H_{c2}^{(1)}/H_{c2}^{(2)}=4.4\pm1.4$] 
is comparable with that of the two energy gaps [$\Delta_1(0)/\Delta_2(0)=4.5(1)$], and that the 
relative weight $w$ also agrees with that deduced from the temperature dependence of 
$\sigma_{\rm v}$ [where $w=0.38(2)$].\cite{Kuroiwa:08} 
Provided that the physical parameters obtained at 2 K is not far 
from those at $T=0$, it might be allowed to discuss the relation between  
the BCS coherence length and energy gap based on the equation,
\begin{equation}
\xi_0=\frac{\hbar v_F}{\pi\Delta(0)}\simeq\xi, \:\:(T\rightarrow0)
\end{equation} 
where $v_F$ is the Fermi velocity.
Since the upper critical field is proportional to the inverse squared of $\xi$
[$H_{c2}=\Phi_0/(2\pi\xi^2)$], one may expect $H_{c2}\propto \Delta^2(0)/v_F^2$.
Thus, the observed coincidence between the ratios
of $H_{c2}$ and of $\Delta(0)$ may imply that the Fermi velocity also varies
between bands nearly by a factor of two. 

Another possible reason for the steeper reduction of superfluid density is the field-induced nodes in the energy gap and associated quasiparticle excitation (QP) specific to noncentrosymmetric 
superconductors.\cite{Fujimoto:07}  The characteristic field, $H_{\rm so}$, may be provided in 
relation to the gap,
\begin{equation}
H_{\rm so}^{(i)}\simeq\frac{\Delta_i(0)}{\mu_B},
\end{equation}
which yields  $H_{\rm so}^{(1)} = 47(2)$ T and  $H_{\rm so}^{(2)} = 10(2)$ T.
It is predicted in the calculation of electronic specific heat that the QP excitation is 
strongly enhanced to reduce the superfluid density towards a field 0.55-0.7$H_{\rm so}$
(depending on the magnitude of the coherence length).  Both of two corresponding
fields, however, seem to be too high to explain the characteristic 
field of kink observed in Fig.~\ref{sgmv} ($\simeq3$ T).  On the other hand,
it is also predicted for the case of small SO interaction [compared with $\Delta(0)$] that the 
QP excitation might be enhanced below $\sim0.1H_{\rm so}$, primarily due to the
asymmetric deformation of the Fermi surface.  In this case, both of the characteristic fields
($\simeq4.7$ T and 1.0 T) are not far from that observed in Fig.~\ref{sgmv}, and 
the weighted average ($\simeq2.5$ T) turns out to be in good accord with the kink of $\sigma_{\rm v}$.

It is presumable as an actual situation that the field dependence of $\sigma_{\rm v}$ reflects 
the effect of double-gap as well as that of the SO interaction, and they are not discernible  
within the present resolution of data along the magnetic field.

In conclusion, we demonstrated the presence of spin-orbit interaction by showing the occurrence of
field-induced enhancement of muon spin relaxation in the normal state of a  noncentrosymmetric superconductor, La$_2$C$_3$.  In the mixed state of La$_2$C$_3$, we also showed that the 
field dependence has a clear kink around 3 T, and that this feature may be well explained by 
considering the effect of i) double-gap structure in the order parameter previously established 
by the temperature dependence of $\sigma_{\rm v}$ and ii) that of the spin-orbit interaction
leading to the field-induced deformation of the Fermi surface.

We would like to thank the staff of TRIUMF for their technical support during the
$\mu$SR experiment.  We also thank S. Fujimoto for helpful discussion on the effect of 
noncentrocymmetry on the electronic property of sesquicarbides.
This work was supported by the KEK-MSL Inter-University Program for Oversea Muon Facilities and 
a Grant-in-Aid for Scientific Research
on Priority Areas by Ministry of Education, Culture, Sports, Science and Technology, Japan.





\end{document}